\documentclass[twocolumn,aps,showpacs,prl,nohypertex,superscriptaddress]{revtex4}
\usepackage{amsmath}
\usepackage{graphicx}
\usepackage{dsfont}
\usepackage{subfigure}
\usepackage{float}
\usepackage{color}
\newenvironment{sequation}{\begin{equation}\small}{\end{equation}}
\begin{document}

\title{Bosonic Integer Quantum Hall State without Landau Levels on Square Lattice}
\author{Wanli Liu}
\thanks{These authors contributed equally to this work.}
\affiliation{Department of Physics and State Key Laboratory of Surface Physics, Fudan University, Shanghai 200433, China}
\author{Zhiyu Dong}
\thanks{These authors contributed equally to this work.}
\affiliation{Department of Physics and State Key Laboratory of Surface Physics, Fudan University, Shanghai 200433, China}
\author{Zhihuan Dong}
\thanks{These authors contributed equally to this work.}
\affiliation{Department of Physics and State Key Laboratory of Surface Physics, Fudan University, Shanghai 200433, China}
\author{Chenrong Liu}
\affiliation{Department of Physics and State Key Laboratory of Surface Physics, Fudan University, Shanghai 200433, China}
\author{Wei Yan}
\affiliation{Department of Physics and State Key Laboratory of Surface Physics, Fudan University, Shanghai 200433, China}
\author{Yan Chen}
\thanks{Corresponding author}
\email{yanchen99@fudan.edu.cn}
\affiliation{Department of Physics and State Key Laboratory of Surface Physics, Fudan University, Shanghai 200433, China}
\affiliation{Collaborative Innovation Center of Advanced Microstructures, Nanjing 210093, China}
\date{\today}

\begin{abstract}
 We study an interacting two-component hard-core bosons on square lattice for which, in the presence of staggered magnetic flux, the ground state is a bosonic integer quantum Hall (BIQH) state. Using a coupled-wire bosonization approach, we analytically show this model exhibits a BIQH state at total charge half filling associated with a symmetry-protected topological phase under $U(1)$ charge conservation. These theoretical expectations are verified, using the infinite density matrix renormalization group method, by providing numerical evidences for: (i) a quantized Hall conductance $\sigma_{xy}=\pm2$, and (ii) two counter-propagating gapless edge modes. Our model is a bosonic cousin of the fermionic Haldane model and serves as an additional case of analogy between bosonic and fermionic quantum Hall states.

 \end{abstract}

\maketitle

\textit{Introduction}.---Topological quantum phases have attracted a tremendous amount of attention in condensed matter physics both theoretically and experimentally during the past three decades~\cite{physics}. They divide into two classes: (i) long-range-entangled phases, characterized by topologically degenerate ground states that support fractionalized excitations with nontrivial braiding statistics, including fractional quantum Hall states\cite{FQH1,FQH2}, spin liquids~\cite{CSL1,CSL2,ZSL1,ZSL2,ZSL3}; and (ii) short-range-entangled phases, protected by a global symmetry~\cite{SPT1,SPT2} and associated with cohomology groups~\cite{group}. Such symmetry-protected topological (SPT) phases have a gapped bulk with no fractionalization and usually still exhibit protected gapless edge excitations~\cite{edge1,edge2,edge3}. The first example of SPT phase is the Haldane spin-1 chain~\cite{spin1,spin2}, which is protected by the spin rotation $SO(3)$ symmetry. Another example is the fermionic topological insulator (TPI)~\cite{TPI1,TPI2,TPI3,TPI4,TPI5}, which is protected by $U(1)$ charge conservation and time-reversal symmetry $T$. Unlike free fermions, where the nontrivial band topology is only required, bosons need strong interaction to stabilize a SPT phase, making them challenging to study.

Recent theoretical studies address the possibility of a bosonic integer quantum Hall (BIQH) state in 2D strongly interacting bosonic systems in the presence of a strong magnetic field. As an analog of the free fermionic TPI, BIQH state is a $U(1)$ SPT phase. It is characterized by an even integer Hall conductance and hosts counter-propagating chiral modes at the edge boundaries~\cite{HE1,HE2}.  According to mutual Chern-Simons theory~\cite{HE1,simons}, BIQH state can be stabilized in two-component bosons in magnetic fields~\cite{two1,two2,two3,two4} at filling factor $\nu\!=\!1$ for each component. The honeycomb lattice model of hard-core bosons with correlated hopping has been studied numerically using the infinite density matrix renormalization group (DMRG)~\cite{BIQH1}, and analytically using the coupled-wire bosonization approach~\cite{BIQH2}. With fine-tuned couplings and internal states in an ultracold atom system, the optical flux lattice in reciprocal space has been studied~\cite{OFL}. More naturally, with Chern bands replacing the lowest Landau level, one may expect the BIQH phases, which has been confirmed in the Harper-Hofstadter model~\cite{Chern} with Chern number $C=2$. In contrast, as an analog of the fermionic Haldane model~\cite{Haldane}, the realization of BIQH state without Landau levels in certain lattice model still remains an open problem.

In this paper, we introduce a simple lattice model of interacting two-component hard-core bosons that realizes BIQH states with no net magnetic field. We first present analytically the possibility of BIQH states using the coupled-wire bosonization method~\cite{HE1,couple}. Next numerical results from DMRG calculations~\cite{IDMRG1,IDMRG2} confirm that its ground state is indeed the BIQH state. In contrast with the BIQH states at filling factors $\nu=1+1$, our lattice model achieves this at total charge half fillings, where the Zeeman energy splitting of the two component bosons is crucial. Finally, a similar lattice model realizing the BIQH state at filling factor $\nu=1+1$ is also introduced and verified numerically.

\begin{figure}
\centering
\includegraphics[width=6.5cm]{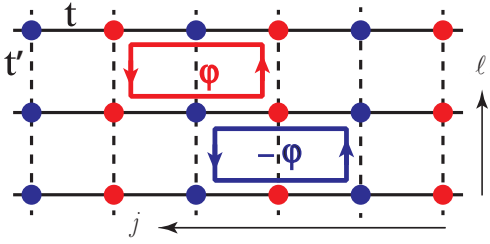}
\caption{(color online). An anisotropic square lattice consisted of A(B) sublattices, with correlated hopping $t$ and normal hopping $t'$ in the two directions. Different species of hard-core bosons feel opposite magnetic flux $\pm\varphi$.}
\label{lattice}
\end{figure}

\textit{Model and coupled-wire bosonization}.---Our model is defined on an anisotropic square lattice (more precisely, weakly coupled wires) with two-component hard-core bosons in the presence of a staggered magnetic flux, which is described by the Hamiltonian $\hat{H}=\hat{H}_0+\hat{H}_1+\hat{H}_{\lambda}$,
\begin{eqnarray}\label{H1}
\nonumber
\hat{H}_0&=&-t\sum_{j,l}
(a^{\dagger}_{j,l}a_{j\!+\!1,l}e^{i\pi n^b_{j,l}}
+b^{\dagger}_{j,l}b_{j\!+\!1,l}e^{i\pi n^a_{j\!+\!1,l}}
+h.c.)\\
\nonumber
&&+\varepsilon\sum_{j,l}
(a_{j,l}^{\dagger}a_{j,l}-b_{j,l}^{\dagger}b_{j,l}),\\
\nonumber
\hat{H}_1&=&t'\sum_{j,l} (a^{\dagger}_{j,l}a_{j,l\!+\!1}e^{i\varphi j}
+b^{\dagger}_{j,l}b_{j,l\!+\!1}e^{-i\varphi j}+h.c.),\\
\hat{H}_{\lambda}&=&\lambda\sum_{j,l}
(a_{j,l}^{\dagger}b_{j,l}+h.c.),
\end{eqnarray}
where $a(b)$ is the annihilation operator for a hard-core boson on sublattice A(B), with $l(j)$ the wire (site) index, as assigned in Fig.~\ref{lattice}. The lattice is treated as coupled wires, with correlated intra-wire hopping~\cite{BIQH1} $t$ and normal inter-wire hopping $t'$. A Zeeman energy splitting $\varepsilon$ and interspecies tunneling $\lambda$ are also considered. Note that in normal hopping the $a$ and $b$ bosons acquire opposite phases ($\varphi=2\pi n_{\varphi}$) through a staggered oscillating flux pattern; see Supplemental Material for more details. The lattice can be realized in ultracold atoms system by trapping different species of atoms at the node or antinode of lasers~\cite{trap}. Alternatively, one can also employ bilayer square lattices. The model breaks time-reversal symmetry $T$ and exhibits a total-charge-conserving $U_c(1)$ symmetry. The correlated hopping $t$ implements mutual flux attachment in the Chern-Simons theory~\cite{simons} and may be realized experimentally by the lattice shaking method~\cite{shaking}.

For simplicity, we first set $\lambda=0$ where a separate number conserving $U(1)\otimes U(1)$ symmetry is featured. Utilizing coupled-wire bosonization, we treat the inter-wire Hamiltonian $H_1$ as a perturbation. To make $H_0$ a simple free Hamiltonian, a Jordan-Wigner-like transformation~\cite{BIQH2} is introduced
\begin{eqnarray}\label{transform}
\nonumber
\tilde{a}_{j,l}&=&O_{j,l}^ba_{j,l},\qquad\qquad\qquad \tilde{b}_{j,l}=O_{j,l}^ab_{j,l}\\
O_{j,l}^a&=&\exp({i\pi\!\sum_{j'>j}\!n_{j',l}^a}),\quad
O_{j,l}^b=\exp({i\pi\!\sum_{j'<j}\!n_{j',l}^b}),
\end{eqnarray}
where $O_{j,l}^{a(b)}$ is the string operator attached to operator $a_{j,l}(b_{j,l})$, one from the left and one from the right. This choice ensures the operators $\tilde{a}$ and $\tilde{b}$ obey bosonic commutation rules. The string operators can also be written in a more symmetric cosine form without changing the physics. The Hamiltonian then becomes
\begin{eqnarray}\label{H2}
\nonumber
\hat{H}_0&=&-t\!\sum_{j,l}(\tilde{a}^{\dagger}_{j,l}
\tilde{a}_{j\!+\!1,l}\!+\!a\leftrightarrow b\!+\!h.c.)+
\varepsilon(n^{\tilde{a}}_{j,l}-n^{\tilde{b}}_{j,l}),\\
\nonumber
\hat{H}_1&=&t'\!\sum_{j,l}
(e^{i\varphi j}\tilde{a}^{\dagger}_{j,l}
\tilde{a}_{j,l\!+\!1}O_{j,l}^bO_{j,l\!+\!1}^b+h.c.)\\
&+&t'\!\sum_{j,l}
(e^{-i\varphi j}\tilde{b}^{\dagger}_{j,l}
\tilde{b}_{j,l\!+\!1}O_{j,l}^aO_{j,l\!+\!1}^a+h.c.).
\end{eqnarray}
Using standard bosonization techniques\cite{bosonization1,bosonization2}, the particle operators ($s=a,b$) are bosonized as
\begin{equation}\label{bosonization}
\tilde{s}_{l}(x)
=\frac{U^se^{-i\sqrt{\pi}\theta_l^s}}{\sqrt{2\pi}}
[1+\cos2\sqrt{\pi}(\phi_l^{s}+\sqrt{\pi}\rho^{s}x)],
\end{equation}
whereas the density operators are bosonized (in the continuum limit with respect to the site index) as $n_l\sim \rho+\partial_x\phi_l/\sqrt{\pi}$. Therefore the Hamiltonian $\hat{H}_0$ of the hard-core bosons is mapped to an array of two-component Luttinger liquids
\begin{equation}\label{HL}
\hat{H}_0=\sum_{l,s=a(b)}\frac{v}{2}\int dx
[(\partial_x\phi_l^s)^2+(\partial_x\theta_l^s)^2],
\end{equation}
where $v=2t\sin(\pi\rho)$ is the velocity of the hard-core boson at the `Fermi level' and $\rho(1\!-\!\rho)$ is the mean density of the $a(b)$ boson. The bosonic fields $\phi_l^s$ and $\theta_l^s$ obey the commutation relations $[\phi_l^s(x),\theta_{l'}^{s'}(x)]=i\delta_{ss'}\delta_{ll'}\Theta(x-x')$, with $\Theta(x-x')$ the Heaviside step function. The Klein factor $U^s$ is chosen as $(-)^{\sum_{(j',l')<(j,l)}N_{j',l'}^s}$ to ensure the boson fields from different wires commute\cite{Klein}. Alternatively, one can map the sublattice to a line and introduce the 2D Jordan-Wigner transformation~\cite{BIQH2,2Dtrans}.

\textit{BIQH states and quantized Hall conductance}.---The BIQH states are realized at integer filling factor $\nu\!=\!\rho/n_{\varphi}\!=\!1$, where the most relevant (scaling dimension is 1) and the nonvanishing terms in $H_1$ are given by
\begin{equation}\label{bH1}
\hat{H}_1=\frac{t'}{\pi}
\!\sum_{l,\mu=1(2)}\!\int\! dx\!\big[\cos\!\sqrt{\pi}
(\tilde{\phi}_l^{\mu}\!+\!\phi_{l\!+\!1}^{\mu})\big].
\end{equation}
Here, the new bosonic fields are introduced
\begin{eqnarray}\label{newb}
\nonumber
\phi_l^1=\theta_l^a\!+\!\phi_l^b
\!+\!\sqrt{\pi}\!N_{l'<l}^a, \quad
\tilde{\phi}_l^1=\phi_l^b\!-\!\theta_l^a
\!+\!\sqrt{\pi}\!N_{l'<l}^a,\\
\phi_l^2=\theta_l^b\!+\!\phi_l^a
\!+\!\sqrt{\pi}\!N_{l'\geq l}^b, \quad
\tilde{\phi}_l^2=\phi_l^a\!-\!\theta_l^b
\!+\!\sqrt{\pi}\!N_{l'\geq l}^b,
\end{eqnarray}
where the notation is abbreviated to $N_{l'<l}^s\equiv \sum_{l'<l}N_{l'}^s$ and $N_{l'\geq l}^s\equiv \sum_{l'\geq l}N_{l'}^s$. These fields satisfy the commutation relations
\begin{eqnarray}\label{commute1}
\nonumber
\big[\partial_x\phi_l^{\mu}(x),\phi_{l'}^{\nu}(x')\big]
&=&2i\delta_{ll'}K_{\mu\nu}\delta(x-x'),\\
\nonumber
\big[\partial_x\tilde{\phi}_l^{\mu}(x),
\tilde{\phi}_{l'}^{\nu}(x')\big]
&=&-2i\delta_{ll'}K_{\mu\nu}\delta(x-x'),\\
\big[\partial_x\phi_l^{\mu}(x),
\tilde{\phi}_{l'}^{\nu}(x')\big]&=&0.
\end{eqnarray}
With the $K$-matrix $K\!\!=\!\!\Big(
     \begin{array}{cc}
        0 & 1 \\
        1 & 0 \\
      \end{array}\Big)$
describing the edge modes in the Chern-Simons theory of the BIQH states~\cite{HE1,simons}. In terms of these new fields, $\hat{H}_0$ is rewritten as
\begin{equation}
\hat{H}_0=\sum_{l,\mu=1(2)}\frac{v}{2}\int dx
[(\partial_x\phi_l^{\mu})^2
+(\partial_x\tilde{\phi}_l^{\mu})^2].
\end{equation}

\begin{figure}
  \centering
  \includegraphics[width=7cm]{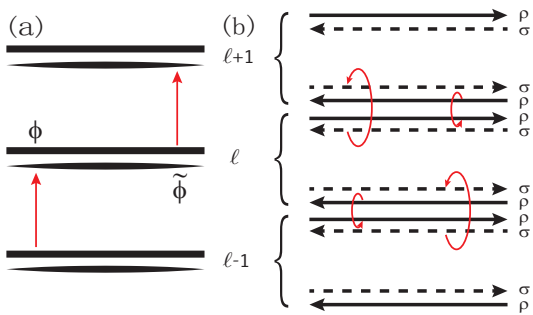}
  \caption{(color online). Schematic of the correlated tunneling processes that lead to the BIQH states. After splitting into charge ($\rho$) and spin ($\sigma$) channels, left (right) movers are coupled with the right (left) movers of neighboring wires separately.}\label{mover}
\end{figure}

In accordance with the renormalization group (RG) flow, the ground state is determined by the perturbation $\hat{H}_1$. Fields $\{\tilde{\phi}_l^1\!+\!\phi_{l\!+\!1}^1\}$ commute with $\{\tilde{\phi}_l^2\!+\!\phi_{l\!+\!1}^2\}$, and hence can be localized simultaneously and gapped out. However, the unpaired edge modes $\phi_1^{\mu}$ and $\tilde{\phi}_N^{\mu}$ remain gapless at the uppermost and lowermost wires; see Fig.~\ref{mover}(a). This is denoted as BIQH$^{+}$ state, with Hall conductance $\sigma_{xy}=2$. Furthermore, reversing the magnetic flux ($\varphi\rightarrow-\varphi$) can be done by shifting the lattice by one grid, and exchanging the particle densities of two species bosons ($\rho\leftrightarrow1\!-\!\rho$) can be done by inverting the energy difference $\varepsilon\rightarrow-\varepsilon$. The most relevant term in $\hat{H}_1$ becomes
\begin{equation}\label{bH2}
\hat{H}_1=\frac{t'}{\pi}\!\sum_{l,\mu=1(2)}\!\int\!dx
\!\big[\cos\!\sqrt{\pi}(\phi_l^{\mu}
\!+\!\tilde{\phi}_{l\!+\!1}^{\mu})\big].
\end{equation}
Now the ground state is characterized by gapless edge modes $\tilde{\phi}_1^{\mu}$ and $\phi_N^{\mu}$, referred to as BIQH$^{-}$ states with Hall conductance $\sigma_{xy}=-2$.

In the composite fermion picture~\cite{composite}, the BIQH$^{+}$ state is realized by attaching one flux quanta to one composite boson whereas the BIQH$^{-}$ state has one hole attached to it. The two BIQH states are dual to each other because of its hard-core nature. The Zeeman energy splitting $\varepsilon$ is crucial as it determines the relative particle density of the different species of bosons in matching the integer filling $\nu=1$, which is approximately expressed as $\varepsilon=2t\cos(\varphi/2)$ when $t'$ and $\lambda$ are very small (see Supplemental Material).

\begin{figure}
\centering
\subfigure{\includegraphics[width=3cm]{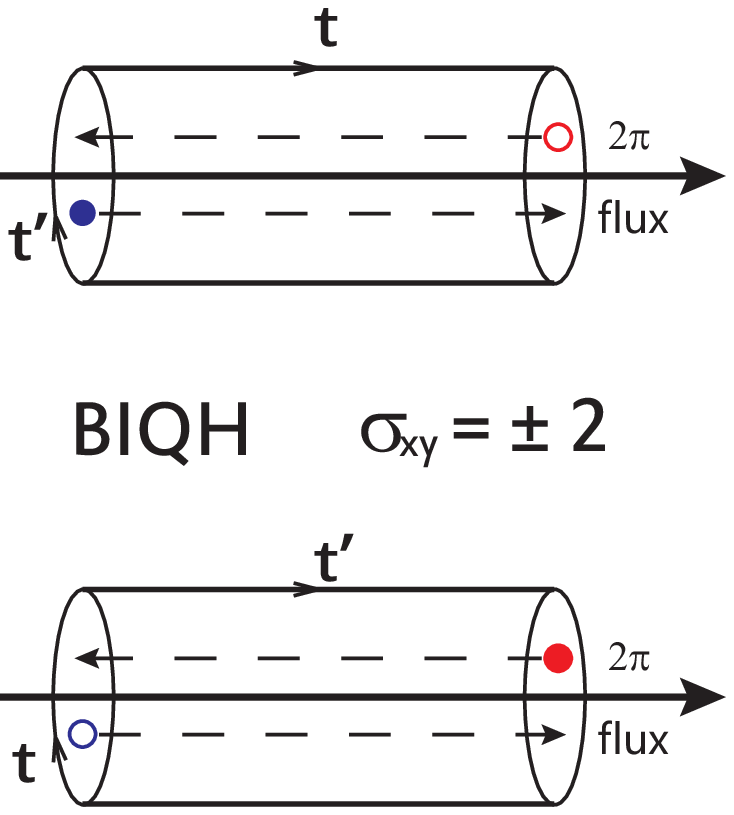}}
\subfigure{\includegraphics[width=5.2cm]{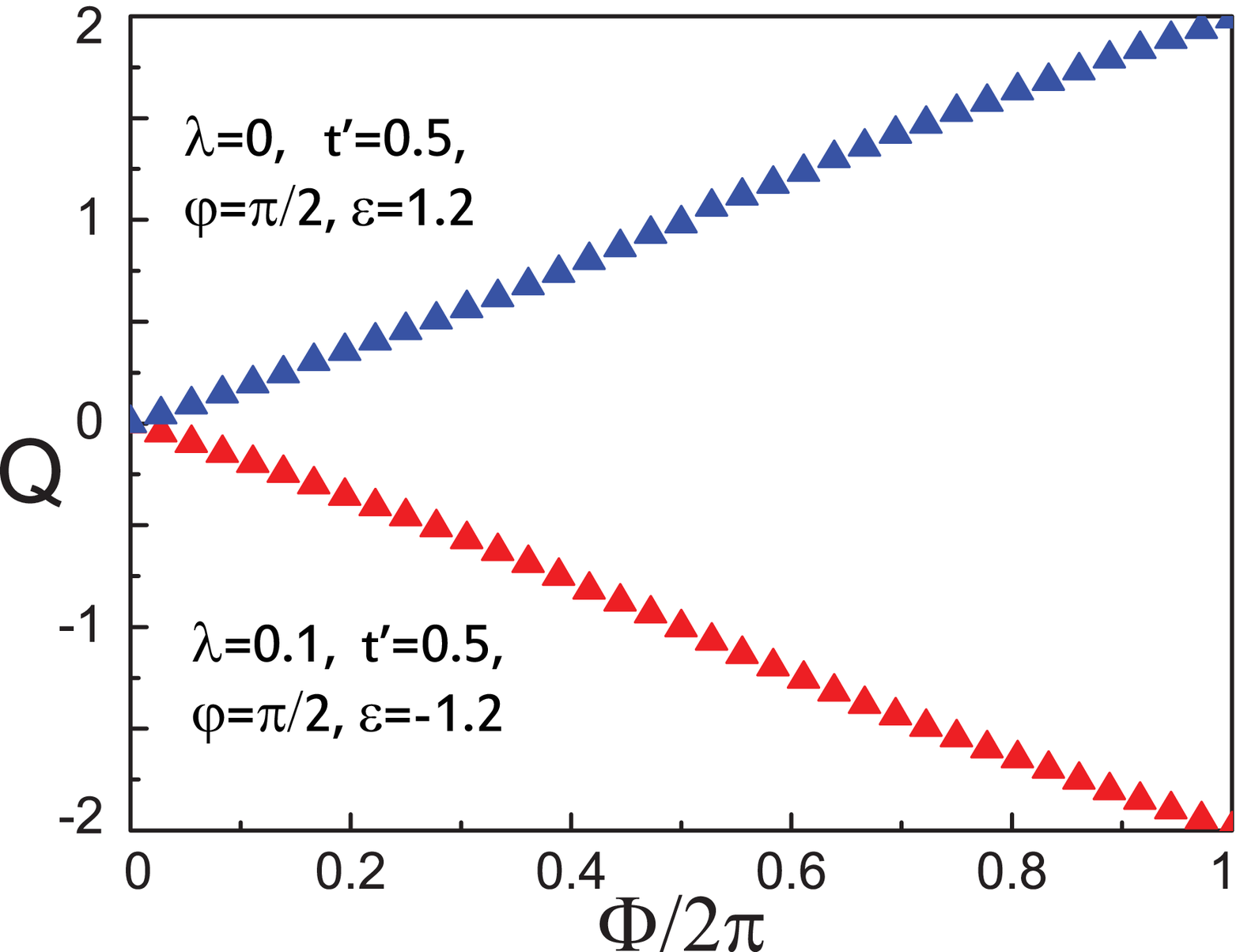}}
\caption{(color online). Charge pumping of the BIQH state with quantized Hall conductance $\sigma_{xy}=\pm2$. Here we show for $t'=0.5t$ and $n_{\varphi}=1/4$, a system with width $L_y=4$. One $a$ particle ($b$ hole) is pumped to the left (right) after a $2\pi$ flux is inserted. For other system sizes or different fluxes, the results are almost the same.}
\label{pump}
\end{figure}

Numerically, we use the infinite DMRG method to study the system wrapped around a cylinder. To measure the Hall conductance $\sigma_{xy}$, we use an adiabatic flux insertion. Specifically, a 2$\pi$ flux insertion pumps the $\sigma_{xy}$ particles from the left edge to the right edge of the cylinder~\cite{pumping1,pumping2}. Based on numerical simulations on twisted lattices along the $t$ and $t'$ directions with wire width of $L_y=4,5,8$, the pumping charge $Q$ clearly features a quantized Hall conductance $\sigma_{xy}=\pm2$ (Fig.~\ref{pump}), which does not depend on the three different geometries. Interestingly, the quantized charge pumping is very robust even when $t'>t$, which is beyond our perturbation description. For the untwisted lattice, the results are almost the same.

\textit{Edge modes and Entanglement spectra}.---From the Chern-Simons theory of BIQH states~\cite{simons}, of which one defining property is the existence of two counter-propagating edge modes, that is, one charge mode carrying charge and one spin mode carrying (pseudo-) spin. Hence the BIQH is a non-chiral phase, contrary to the TPI with only one chiral mode. To explore charge and spin channel more explicitly, it is useful to define left and right moving chiral field $\phi^{L/R}=(\phi\pm\theta)$ as well as charge and spin fields $\phi_{\rho/\sigma}^p=(\phi_a^p\pm\phi_b^p)/\sqrt{2}$ for $p=R,L$. These chiral fields defined on the wires satisfy commutation relations
\begin{equation}\label{commute2}
\big[\partial_x\phi_{l,\mu}^p(x),
\phi_{l',\mu'}^{p'}(x')\big]
=2ip\delta_{ll'}\delta_{\mu\mu'}\delta_{pp'}\delta(x-x'),
\end{equation}
where $p=\pm1$ for R(L)-movers and $\mu=\rho,\sigma$. In terms of these fields, $H_0$ is written as
\begin{equation}\label{chiralH0}
\hat{H}_0=\sum_{l,p,\mu}\frac{v}{4}\int dx
(\partial_x\phi_{l,\mu}^p)^2.
\end{equation}
We expect $v_c=v_s$ when $\lambda=0$ because of the symmetry of the Hamiltonian. In the BIQH$^+$ phase, the inter-wire Hamiltonian $H_1$ becomes
\begin{sequation}
\hat{H}_1\sim\!\sum_l\!\int\! dx
\cos\!\sqrt{\frac{\pi}{2}}
(\phi_{l,\rho}^R\!+\!\phi_{l\!+\!1,\rho}^L)
\cos\!\sqrt{\frac{\pi}{2}}
(\phi_{l,\sigma}^L\!+\!\phi_{l\!+\!1,\sigma}^R).
\end{sequation}
One finds that right (left) movers are coupled with left (right) movers of the neighboring wires in each channel, whereas  $(\phi_{1,\rho}^L,\phi_{1,\sigma}^R)$ and $(\phi_{N,\rho}^R,\phi_{N,\sigma}^L)$ are left gapless, indicating two counter-propagating chiral edge modes; see Fig.~\ref{mover}(b). The situation for the BIQH$^-$ phase is similar.

The corresponding effective edge Hamiltonian is\cite{two1}
\begin{equation}\label{edgeH}
\hat{H}=\frac{2\pi}{L_y}(v_c\hat{L}_0^c+v_s\hat{L}_0^s),
\quad \hat{P}=\frac{2\pi}{L_y}(\hat{L}_0^c-\hat{L}_0^s),
\end{equation}
with
\begin{equation}\label{momentum}
\hat{L}_0^{c(s)}
=\frac{(\triangle N_a\pm\triangle N_b)^2}{4}
+\sum_{m=1}^{\infty}m\hat{n}_m^{c(s)}.
\end{equation}
Here $L_y$ is the length of the 1D edge, $\triangle N_{a(b)}$ is the change in the $a(b)$ particle number relative to the ground state, and $\{n_m^{c(s)}\}$ is the number of oscillator modes with momentum number $m$. These bosonic oscillator modes exhibit the well-known $1,1,2,3,\cdots$ degeneracy pattern.
\begin{table}
  \centering
  \caption{Energies and degeneracies of different states of the entanglement Hamiltonian (edge mode) in different charge sectors $\triangle N_a+\triangle N_b=n (n=0,\pm1,\pm2...)$}.\label{degeneracy}
  \includegraphics[width=8.4cm]{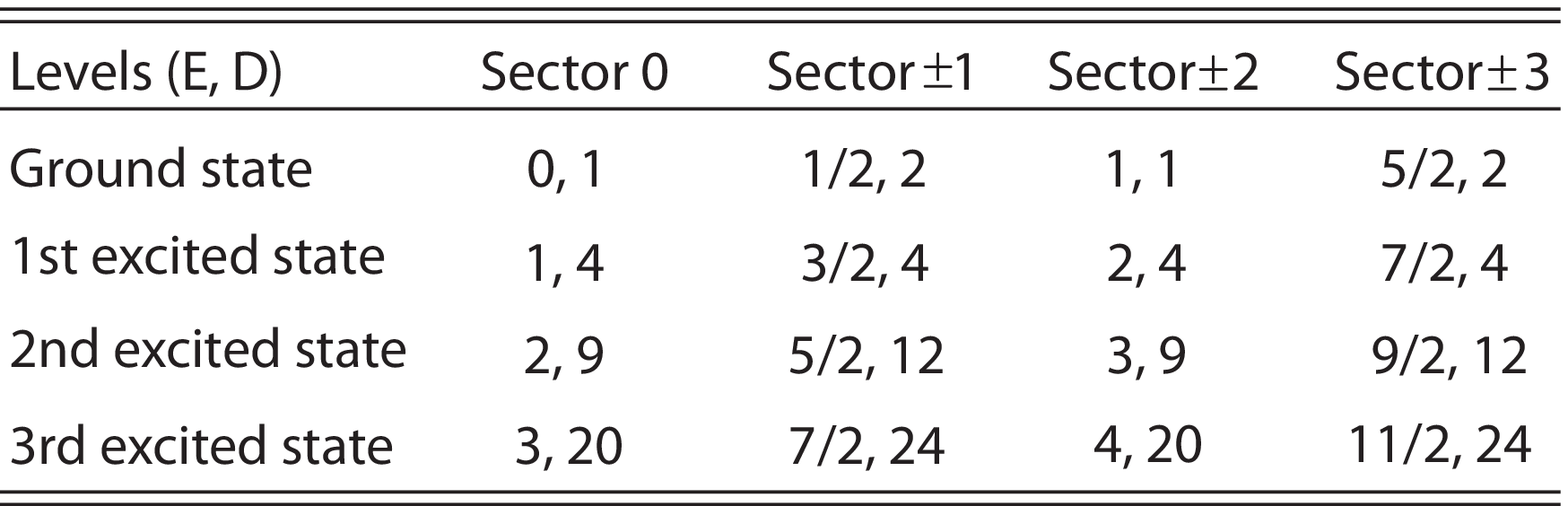}
\end{table}

\begin{figure}
\centering
  \includegraphics[width=7.5cm]{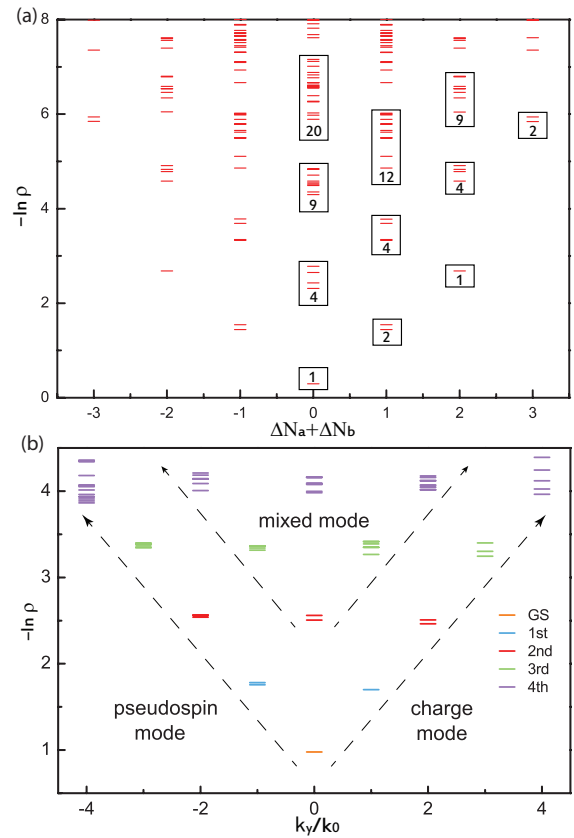}
  \caption{(color online). (a) Entanglement spectra in different charge sectors, with numbers indicating degeneracies of different energy levels. (b) Entanglement spectra versus momentum in charge sectors $\Delta N_a+\Delta N_b=0$, with dashed lines denoting the edge modes. Here $k_0=2\pi/L_y$ is the quanta of momentum. Numerical simulations are based on an infinite cylinder of width $L_y=4$, $\lambda=0$, $n_{\varphi}=1/4$, and $t'=0.5 (a)$, $t'=0.2 (b)$.}
\label{entangle}
\end{figure}

When $\lambda=0$, the energy level for the edge Hamiltonian is determined by quantum number $\hat{L}_0^c+\hat{L}_0^s$. Degeneracies come from the different distributions of $L_0^{c(s)}$ as well as number $\triangle N_{a(b)}$ and oscillating quantum number $\{n_m^{c(s)}\}$, which can be calculated straightforwardly. More details are elaborated in Table I and Supplemental Material.

Numerically, we use the entanglement spectra as a probe of the edge modes~\cite{ES} because the low-energy spectra of the system in the presence of a physical edge coincides with the low-lying part of the entanglement spectra, up to rescaling and shifting. Numerical results from the DMRG simulation are shown in Fig.~\ref{entangle}. We have plotted several different cases that correspond to the $U(1)$ charge sector $\triangle N_a+\triangle N_b=n(n=0,\pm1,\pm2,\cdots)$. The energies and degeneracies in each sector are in perfect accord with the theoretical expectations (Table \ref{degeneracy}). In particular, we plot the entanglement spectra versus momentum in the charge sector $\triangle N_a+\triangle N_b=0$. The edge modes are linear with momentum $k$, whereas the charge sector shows a $1,2,3,\cdots$ degeneracy pattern in the charge channel and a $3,4,7,\cdots$ pattern in the pseudospin channel~\cite{BIQH1}.

\textit{Effect of interspecies tunneling}---When turn on the interspecies tunneling $\hat{H}_{\lambda}$, the $U(1)\otimes U(1)$ symmetry now breaks down to a global $U_c(1)$ symmetry. We find the quantized charge pumping and counter-propagating edge modes still robust for a finite $\lambda(\leq 0.6)$, and almost all the properties are the same as for $\lambda=0$. Here $\hat{H}_{\lambda}$ is defined on intra-wires and gives rise to coupling terms
\begin{equation}
\hat{H}_{\lambda}\sim\lambda\sum_{l}\!\int\!dx
\big(\cos\!\sqrt{2\pi}\phi_{l,\sigma}^R
+\cos\!\sqrt{2\pi}\phi_{l,\sigma}^L\big).
\end{equation}
At first sight they appear to open gaps for the (pseudo-) spin edge modes, but actually they do not satisfy Haldane¡¯s null-vector criterion~\cite{criterion} and the two cosine terms cannot be localized because they are chiral and do not commute with each other. In general, the counter-propagating edge modes are robust against perturbations~\cite{edge2,HE1,simons,BIQH2}, and interspecies tunneling does not modify the qualitative nature of the BIQH states, as long as the $U_c(1)$ symmetry is conserved.

\textit{BIQH states at $\nu=1\!+\!1$}.---Apart from the BIQH states with no net magnetic field at total charge half filling, the BIQH states can also be realized at filling factor $v=1$ for each component in a similar model. The staggered magnetic flux is now replaced by a uniform flux (the $a$ and $b$ bosons acquire the same phases in normal tunneling), and the Zeeman energy splitting $\varepsilon$ is set to zero. It is easy to show in a similar manner that this model supports BIQH states. Its quantized charge pumping (see Supplemental Material) and entanglement spectra are almost the same as the case without Landau levels.

\textit{Conclusion}---We have introduced a simple interacting two-component bosonic lattice model on square lattice that realizes a BIQH state with no net magnetic flux. This state is realized at total charge half filling and protected by a global $U_c$ symmetry. The Zeeman energy splitting is crucial to match the integer filling $\nu=1$ for particles or holes. Using the coupled-wire bosonization techniques as well as numerical DMRG methods, we demonstrated the two defining characteristics of the BIQH states: (i) quantized Hall conductance $\sigma_{xy}=\pm2$ and (ii) two counter-propagating gapless edge modes. This model is a bosonic cousin of the fermionic Haldane model~\cite{Haldane}, and serves as an additional case of analogy between fermionic and bosonic quantum Hall states. Finally, with a similar model, we introduced a realization of the BIQH states at filling $\nu=1+1$.

\textbf{Acknowledgements.} We acknowledge Y.-C. He for fruitful discussions and introducing to this problem. We thank Y.S. Wu, R.B. Tao, T.K. Lee, and Z.D. Wang for helpful discussions. This work was supported by the State Key Programs of China (Grant No. 2017YFA0304204£¬and 2016YFA0300504), the National Natural Science Foundation of China (Grant Nos. 11625416, and 11474064).

\end{document}